\documentclass[journal,10pt,twocolumn,twoside]{IEEEtran}

\ifCLASSINFOpdf
\else
\fi
\usepackage[cmex10]{amsmath}
\usepackage{amssymb}
\usepackage{color}
\usepackage{tikz}
\usetikzlibrary{decorations.pathmorphing}
\usetikzlibrary{fit,shapes}
\tikzstyle{var}=[circle,thick,draw=black,minimum size=1.2cm]
\usepackage{subfig}

\newtheorem{theorem}{Theorem}

\newtheorem{lemma}[theorem]{Lemma}

\newcommand{\PP}{\mathbb{P}}
\newcommand{\todo}[1]{{#1}}
\newcommand{\rev}[1]{{#1}}
\newcommand{\greg}[1]{{#1}}
\newcommand{\change}[1]{{#1}}

\begin{document}
%
\title{Measuring the Influence of Observations in HMMs through the Kullback-Leibler Distance}
%
%
%

\author{Vittorio~Perduca
        and~Gregory~Nuel
\thanks{V. Perduca and G. Nuel are with the Laboratoire MAP5, Universit\'e Paris Descartes and CNRS, Sorbonne Paris Cit\'e, France, email: \texttt{\{vittorio.perduca, gregory.nuel\}@parisdescartes.fr}.}}

\maketitle

\begin{abstract}
We measure the influence of individual observations on the sequence of the hidden states of the Hidden Markov Model (HMM) by means of the Kullback-Leibler distance (KLD). Namely, we consider the KLD between the conditional distribution of the hidden states' chain given the complete sequence of observations and the conditional distribution of the hidden chain given all the observations but the one under consideration. We introduce a linear complexity algorithm for computing the influence of all the observations. As an illustration, we investigate the application of our algorithm to the problem of detecting {meaningful observations} in HMM data series.
\end{abstract}

\begin{IEEEkeywords}
Hidden Markov models, relative entropy, forward-backward algorithm, outlier detection, local outlier factor
\end{IEEEkeywords}

%
\IEEEpeerreviewmaketitle

\section{Introduction}
\IEEEPARstart{T}{he} Hidden Markov Model (HMM) is a standard tool in many applications, including signal processing and speech recognition \cite{ephraim2002hidden,rabiner1989tutorial,gold2011speech} and computational biology \cite{Durbin1999}. In a typical HMM, let $S_{1:n}=(S_1,\ldots,S_n)$ be the Markov sequence of hidden variables (or states) and $X_{1:n}=(X_1,\ldots,X_n)$ the sequence of observation variables\footnote{In the absence of a widespread standard notation, symbols denoting HMM variables vary from author to author.}. In this letter we address the problem of measuring the influence of an observation $X_j=x_j$ on the distribution of the hidden sequence $S_{1:n}$. 

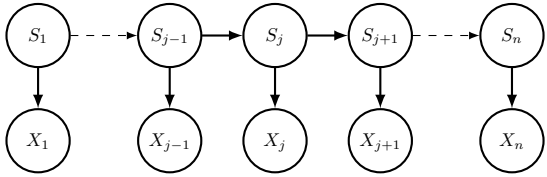
\begin{figure}[t]
\begin{center}
\begin{tikzpicture}[>=latex,text height=1.5ex,text depth=0.25ex,scale=0.7, transform shape] 
   \draw (-0.5,0) node (X1) [var] {$X_1$};
   \draw (2,0) node (X2) [var] {$X_{j-1}$};
   \draw (4,0) node (X3) [var] {$X_{j}$};
   \draw (6,0) node (X4) [var] {$X_{j+1}$};
   \draw (8.5,0) node (X5) [var] {$X_n$};
   \draw (-0.5,2) node (S1) [var] {$S_1$};
   \draw (2,2) node (S2) [var] {$S_{j-1}$};
   \draw (4,2) node (S3) [var] {$S_j$};
   \draw (6,2) node (S4) [var] {$S_{j+1}$};
   \draw (8.5,2) node (S5) [var] {$S_n$};
   \path[->]
       (S1) edge[thick] (X1)
       (S2) edge[thick] (X2)
       (S3) edge[thick] (X3)
       (S4) edge[thick] (X4)
       (S5) edge[thick] (X5)
       (S1) edge[dashed] (S2)
       (S2) edge[thick] (S3)
       (S3) edge[thick] (S4)
       (S4) edge[dashed] (S5)
       ;
\end{tikzpicture}
\end{center}
\caption{HMM topology. $S_j$: hidden variable, $X_j$: observed variable.}
\label{fig:hmm}
\end{figure}

We start by fixing notation. For simplicity's sake, we consider \emph{homogeneous} HMMs and denote the parameters of the model with $\PP(X_i=x|S_i=s)=\beta(s,x)$ (\emph{emissions}), $\PP(S_i=s|S_{i-1}=r)=\alpha(r,s)$ (\emph{transitions}) and $\PP(S_1=s)=\gamma(s)$. The model is fully specified by the conditional dependencies among the variables depicted in Fig. \ref{fig:hmm} which determine the following factorization of the joint probability distribution 
\begin{multline*}
\PP(X_{1:n}=x_{1:n},S_{1:n}=s_{1:n})=\\
\gamma(s)\prod_{i=2}^n\alpha(s_{i-1},s_i)\prod_{i=1}^n\beta(s_i,x_i),
\end{multline*}
where $s_i,x_i$ are taken in the sets of all possible outcomes of $S_i$ and $X_i$ (for continuous variables simply replace probabilities with densities and sums with integrals). For simplicity of notation, in most equations we omit to write explicitly the outcomes of the variables.  

An important inference problem in HMMs is computing the conditional (posterior) distribution of the hidden sequence given an \emph{evidence}. In standard applications, the evidence is a complete instantiation of the observable sequence, $\mathcal{E}=\{X_{1:n}=x_{1:n}\}$ for some $x_{1:n}$. For a fixed $j\in\{1,\ldots,n\}$, we denote $\mathcal{E}_{-j}$ the evidence $\{X_{-j}=x_{-j}\}$, where $X_{-j}$ denotes the sequence of all the observation variables except $X_j$. 

Our suggestion for measuring the influence of an observation $X_j=x_j$ is based on the following question: what is the contribution of $X_j=x_j$ to the posterior distribution $S_{1:n}|\{X_{1:n}=x_{1:n}\}$ of the hidden sequence given the complete sequence of observations? That is, how \emph{dissimilar} are the posterior distributions $\PP(S_{1:n}|\mathcal{E})$ and $\PP(S_{1:n}|\mathcal{E}_{-j})$? The more these two posterior distributions are \emph{distant}, the more $X_j=x_j$ must be influential.

The Kullback-Leibler distance (KLD) (or relative entropy) arises in many applications as an appropriate measurement of the distance between two probability distributions \cite{Bishop2006,do2003fast}. Following \cite{cook1977detection} and \cite{johnson1985influence} (in the context of linear regression), we suggest to measure the influence of $X_j=x_j$ through the KLD  
\begin{equation*}
\label{eq:kld}
 K_j := \sum_{S_{1:n}}\PP(S_{1:n}|\mathcal{E}_{-j}) \log\frac{\PP(S_{1:n}|\mathcal{E}_{-j})}{\PP(S_{1:n}|\mathcal{E})}.
\end{equation*}
\greg{By definition, $K_j$ measures the influence of observation $x_j$ on the posterior distribution of the hidden states rather than on the parameter estimate. $K_j$ is therefore an appropriate influence measure when the quantity of interest is the posterior distribution as it is often the case in practical applications such as speech recognition \cite{gold2011speech}, data segmentation {\cite{fridlyand04}}, bioinformatics {\cite{camproux2004hidden}}, genetics {\cite{li2010mach}}.}

In this letter we address the problem of computing efficiently the vector $(K_j)_{j=1,\ldots,n}$ of all the KL distances, one for each observation. To the best of our knowledge, the computation of the KLD between the posterior distributions of the hidden sequence of an HMM conditioned on two distinct evidences was not studied before, the main efforts being rather aimed at computing efficiently the KLD between the distributions of the observation sequence of an HMM with respect to two distinct sets of parameters \cite{do2003fast, silva2008upper, sahraeian2011novel}. 

A straightforward computation of $(K_j)_{j=1,\ldots,n}$ based on the standard forward-backward algorithm for HMMs leads to a quadratic complexity in the number of observations; our main contribution is a linear time algorithm based on simple recursive formulae. 

\greg{As an illustration, we apply our algorithm to a time series of temperature changes and discuss the practical interest of the suggested influence measure.}


\section{Computation of the Influence Measure}

We start by recalling that the posterior distribution $\PP(S_{1:n}|\mathcal{E})$ of the hidden sequence given the standard evidence is an heterogeneous Markov sequence whose transition probabilities are computed in $\mathcal{O}(nm^2)$ steps with the standard forward-backward algorithm, where $m$ is the is the number of possible outcomes of each hidden variable, \cite{rabiner1989tutorial}. \rev{The forward and backward quantities are defined as $F_i(s):=\PP(X_{1:i},S_i=s)$, $B_i(s)=\PP(X_{i+1:n}|S_i=s)$ and are computed recursively with Eqs. (\ref{eq:forw}) and (\ref{eq:back}).}

Similarly to $\PP(S_{1:n}|\mathcal{E})$, the computation of $\PP(S_{1:n}|\mathcal{E}_{-j})$, for a fixed $j$, requires $\mathcal{O}(nm^2)$ steps: it is easy to adapt the forward-backward algorithm to $\mathcal{E}_{-j}$ by simply marginalizing out the variable $X_j$ in all the formulae and propagating the new forward and backward quantities thus obtained. 

It is straightforward to compute recursively the KLD between two heterogeneous Markov Chain;  as a consequence, a direct approach based on standard recursions leads to a $\mathcal{O}(nm^2)$ time complexity for computing $K_j$ for a fixed $j$. However, the required marginalization of the forward and backward quantities depends on the fixed $j$ and therefore it is necessary to compute a distinct set of forward and backward quantities for each $j$. As a consequence, the resulting complexity for computing the vector $(K_j)_{j=1,\ldots,n}$ is $\mathcal{O}(n^2m^2)$. 

Our principal contribution are new recursive formulae that reduce this complexity to $\mathcal{O}(nm^2)$. We start with two technical lemmas that lead to our original algorithm. 

\begin{lemma} 
For an arbitrary fixed $j\in\{1,\ldots,n\}:$
\begin{equation*}
\label{eq:kld_j}
K_j =  \sum_{S_j}\PP(S_j|\mathcal{E}_{-j})\log\frac{\PP(S_j|\mathcal{E}_{-j})}{\PP(S_j|\mathcal{E})}.
\end{equation*}
\end{lemma}

\begin{IEEEproof}
We start by observing that the following factorizations hold:
$$
\PP(S_{1:n}|\mathcal{E}_{-j})=\PP(S_j|\mathcal{E}_{-j})\PP(S_{1:j-1}|S_j,\mathcal{E}_{-j})\PP(S_{j+1:n}|S_j,\mathcal{E}_{-j}) 
$$
and
$$
\PP(S_{1:n}|\mathcal{E})=\PP(S_j|\mathcal{E})\PP(S_{1:j-1}|S_j,\mathcal{E})\PP(S_{j+1:n}|S_j,\mathcal{E}).
$$
The key point is that in the last equation we have 
$$
\PP(S_{j+1:n}|S_j,\mathcal{E}) = \PP(S_{j+1:n}|S_j,\mathcal{E}_{-j})
$$
$$
\PP(S_{1:j-1}|S_j,\mathcal{E})=\PP(S_{1:j-1}|S_j,\mathcal{E}_{-j})
$$
because $S_{j+1:n}$ and $X_j$ are conditionally independent given $S_j$, and $S_{1:j-1}$ and $X_j$ are conditionally independent given $S_j$, see Fig. \ref{fig:hmm}. \rev{Then $K_j=\sum_{S_j}\PP(S_j|\mathcal{E}_{-j})\log\frac{\PP(S_j|\mathcal{E}_{-j})}{\PP(S_j|\mathcal{E})}\times$ \\
$\sum_{S_{1:j-1}}\PP(S_{1:j-1}|S_j,\mathcal{E}_{-j})\sum_{S_{j+1:n}}\PP(S_{j+1:n}|S_j,\mathcal{E}_{-j})$.   
}
\end{IEEEproof}

As a consequence of this lemma, the key for computing efficiently $(K_j)_{j=1,\ldots,n}$ is an efficient computation of the factors $\PP(S_j|\mathcal{E})$ and $\PP(S_j|\mathcal{E}_{-j})$ for all $j=1,\ldots,n$. For a given $j$, $\PP(S_j|\mathcal{E})$ can be computed in $\mathcal{O}(nm^2)$ steps using the standard forward-backward algorithm: $\PP(S_j=s,\mathcal{E})=F_j(s)B_j(s)$ and hence $\PP(S_j=s|\mathcal{E})\propto F_j(s)B_j(s)$, where the standard forward and backward quantities are computed recursively with 
\begin{equation}
\label{eq:forw}
F_i(s)=\sum_rF_{i-1}(r)\alpha(r,s)\beta(s,x_i)
\end{equation}
and
\begin{equation}
\label{eq:back}
B_{i-1}(r) = \sum_s \alpha(r,s)\beta(s,x_i)B_i(s).
\end{equation} 
We show a similar result for $\PP(S_{j}|\mathcal{E}_{-j})$:

\begin{lemma} For an arbitrary fixed $j\in\{1,\ldots,n\}:$
$$
\PP(S_j=s,\mathcal{E}_{-j})= F^*_j(s)B_j(s),
$$
where $B_j$ is the standard backward quantity and $F^*_j$ is computed recursively from the standard forward quantities with
\begin{equation}
\label{eq:forw_star}
F^*_i(s)=\sum_r F_{i-1}(r)\alpha(r,s) \mbox{ for } i=2,\ldots,n,
\end{equation}
with $F^*_1(s)=\gamma(s)$. Moreover the time complexity for computing $\PP(S_j=s|\mathcal{E}_{-j})\propto F^*_j(s)B_j(s)$ for all $j=1,\ldots,n$ is $\mathcal{O}(nm^2)$. 
\end{lemma}

\begin{IEEEproof}
For a given $j$, we have $\PP(S_j=s,\mathcal{E}_{-j})=$
$$
\sum_{y}\PP(S_j=s,\mathcal{E}_{-j},X_j=y)=\sum_{y}\PP(S_j=s,\mathcal{E}_{y}),
$$
where $\mathcal{E}_{y}$ is the standard evidence $\{X_{-j}=x_{-j},X_j=y\}$. For each $y$ there is a distinct set of standard forward and backward quantities $F_i^{y}, B_i^{y}$; however it is easy to see that $F_i^{y}\equiv F_i$ for $i\leq j-1$ and $B_i^{y}\equiv B_i$ for $i\geq j$. It follows that $\sum_{y} \PP(S_j=s,\mathcal{E}_{y})=$
\begin{multline*}
\sum_{y}F_j^{y}(s)B_j^{y}(s)=B_j(s)\sum_{y}F_{j}^{y}(s)=\\
B_j(s)\sum_{y}\sum_{r}F_{j-1}^{y}(r)\alpha(r,s)\beta(s,y)= \\
B_j(s)\sum_{r}F_{j-1}(r)\alpha(r,s).
\end{multline*}
\end{IEEEproof}

Our main result is a straightforward consequence of the two lemmas above:

\begin{theorem} For an arbitrary fixed $j\in\{1,\ldots,n\}:$ 
$$
K_j=\sum_s \frac{F^*_j(s)B_j(s)}{\sum_rF^*_j(r)B_j(r)}\log\left( \frac{F_j^*(s)}{F_j(s)}\cdot\frac{\sum_r F_j(r)B_j(r)}{\sum_r F_j^*(r)B_j(r)}\right),
$$
where the quantities $(F_i)_{i=1,\ldots,n},(B_i)_{i=1,\ldots,n}$ and $(F^*_i)_{i=1,\ldots,n}$ are computed once and for all independently of $j$ using the recursions (\ref{eq:forw}), (\ref{eq:back}), (\ref{eq:forw_star}). The complexity of computing $(K_j)_{j=1,\ldots,n}$ is $\mathcal{O}(nm^2)$.
\end{theorem}


\section{\change{Application to Time Series Segmentation}}

\todo{We illustrate the practical interest of our influence measure on a real dataset, namely}
\greg{a time series consisting of 106 annual changes in global temperature between 1880 and 1985 \cite{data}. Following the approach suggested by \cite{fridlyand04}, the dataset can be modeled with an homoscedastic HMM in which each observation follows a Gaussian distribution whose mean depends on the corresponding hidden state; we assume that there are three hidden states. We estimated the parameters of the HMM with the EM algorithm and obtained \todo{for the three hidden Gaussian distributions the means} $\mu_1=-0.372$, $\mu_2=0.069$, $\mu_3=-0.068$, and standard deviation $\sigma=0.114$; \todo{moreover the transition matrix is $\pi(i,j)=\eta/2$ if $i\neq j$ and $\pi(i,i)=1-\eta$, where the estimated transition rate is $\eta=0.085$.}} 

\begin{figure}[!t]
\centering
\includegraphics[width=0.45\textwidth,trim=0 260 0 0,clip]{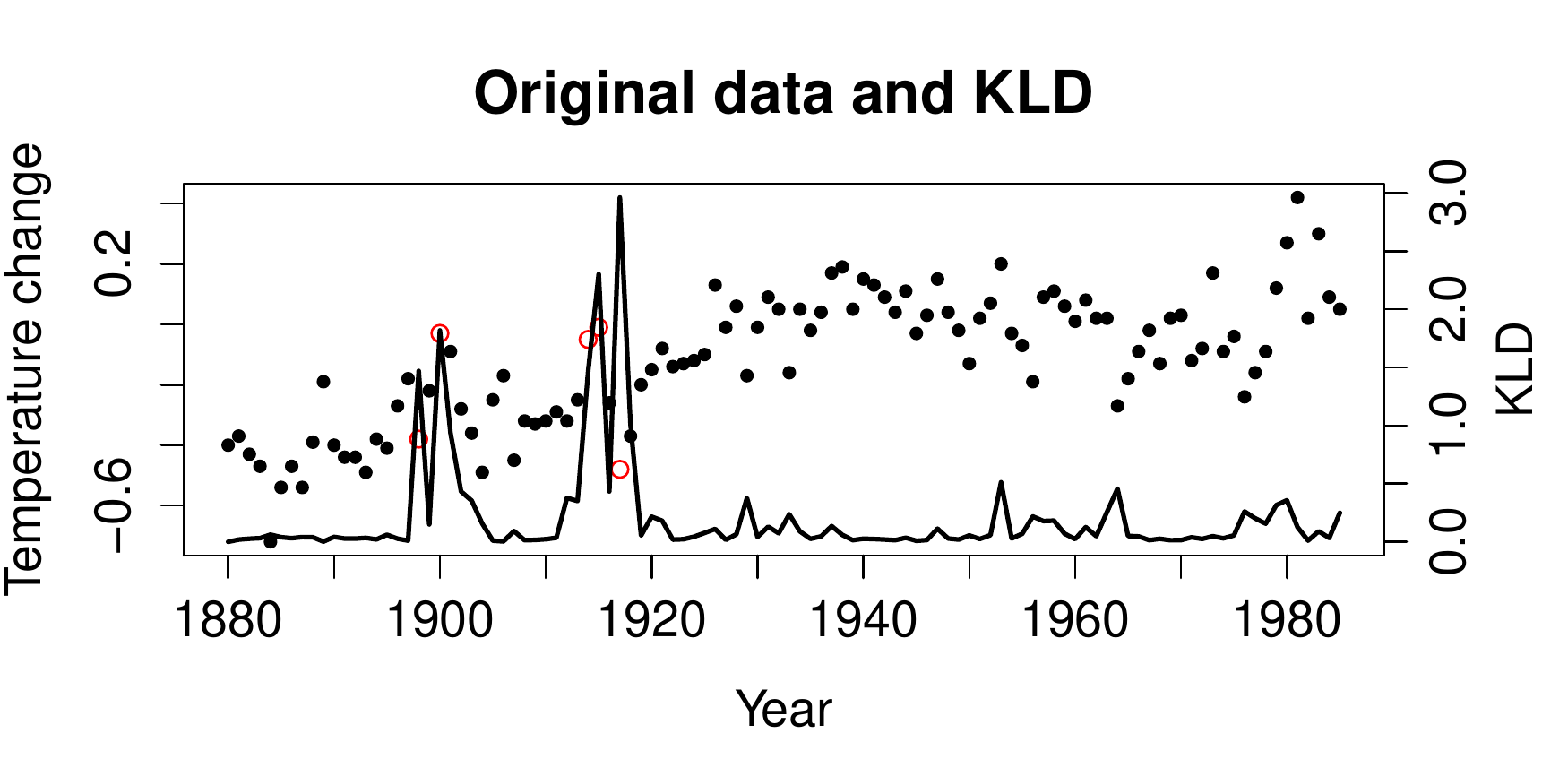}
\caption{\greg{KLD function $(K_{j})_{j=1,\ldots,n}$ for the temperature change time series. The five highest $K_j$ \todo{are}: $K_{1917}=2.96$, $K_{1915}=2.30$, $K_{1900}=1.82$, $K_{1898}=1.47$, $K_{1914}=1.46$; \todo{the corresponding datapoints are depicted as empty dots}.}}
\label{fig:data_kld}
\end{figure}

\greg{Fig.~\ref{fig:data_kld} shows the temperature time series together with the KLD $K_j$ for each $j$. Five years clearly appear to have a greater influence on the posterior distribution of the hidden states: 1917, 1915, 1900, 1898, 1914. \todo{It might be interesting to investigate the reasons why these five years are so influential, looking for either specific climatic events or possible changes in the data collection protocol.}}

\begin{figure}[!t]
\centering
\includegraphics[width=0.45\textwidth]{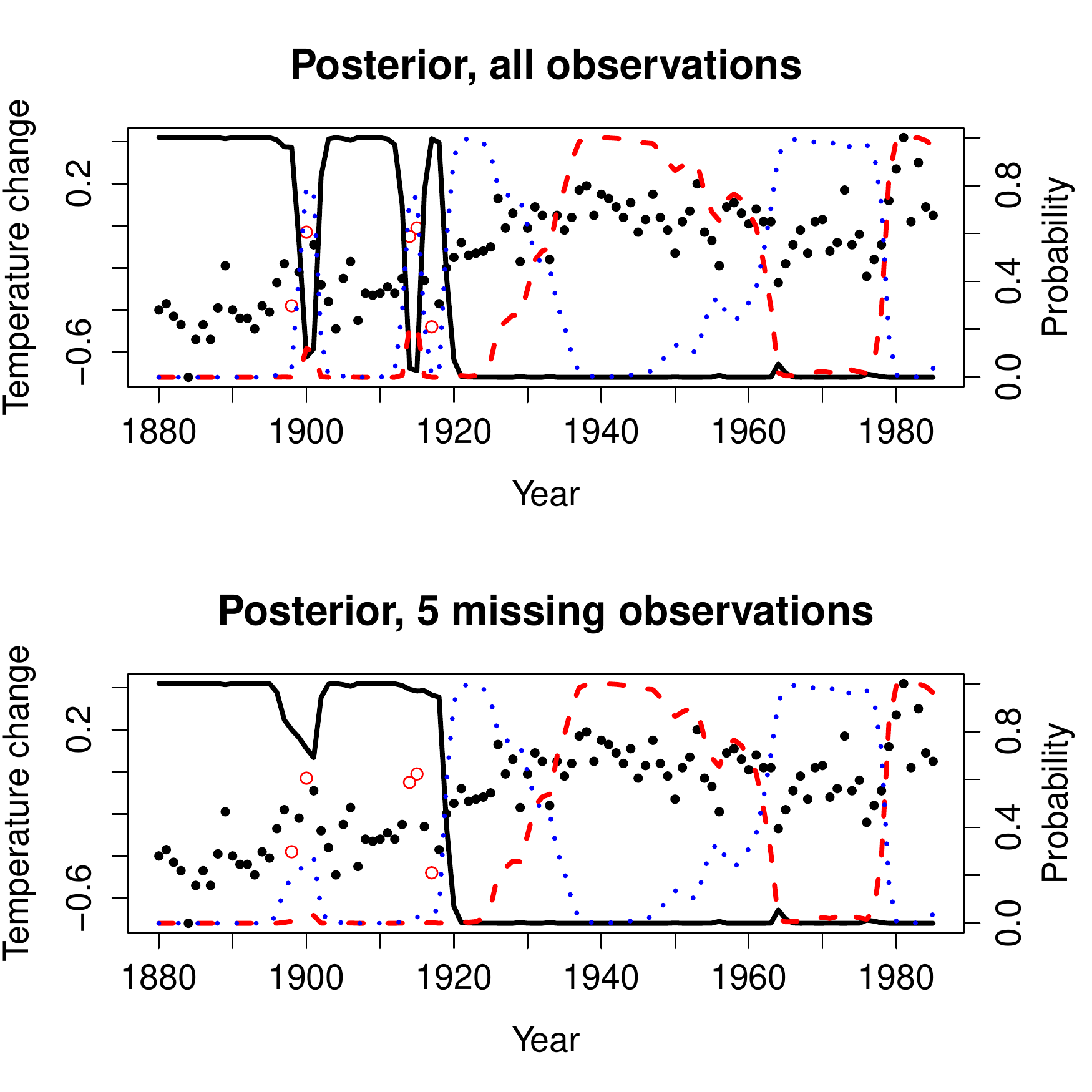}
\caption{\greg{Marginal posterior distributions \todo{$\PP(S_j|\mbox{obs})$} of the three-level segmentation considering all the observations (top) and after removing the five most influential ones (bottom). \todo{Solid black line: $\PP(S_j=1|\mbox{obs})$, i.e. $x_j$ is Gaussian with mean $\mu_1=-0.372$; dashed red: $\PP(S_j=2|\mbox{obs})$ with $\mu_2=0.069$;  dotted blue: $\PP(S_j=3|\mbox{obs})$ with $\mu_3=-0.068$.}}}
\label{fig:data_post}
\end{figure}

\greg{\todo{In order to validate the findings in Fig.~\ref{fig:data_kld}}, we further investigated the effect of the five most influential observations on the posterior distribution of the segmentation by comparing the marginal posterior distributions obtained with all the observations and after removing the five most influential ones, see Fig.~\ref{fig:data_post}. Unsurprisingly, the most dramatic changes occur in the neighborhood of the removed data. \todo{When all the observations are taken into account}, the period 1880-1920 is \todo{characterized} by a long segment of negative annual temperature change interrupted by two short periods of slightly positive annual change around years 1900 and 1914 (Fig.~\ref{fig:data_post}, top). \todo{When the most influential observations are not considered}, these two interruptions basically disappear (Fig.~\ref{fig:data_post}, bottom).}

\greg{The KLD measure of influence is hence clearly \todo{effective in pointing out} observations that have a dramatic effect on the posterior segmentation. These observations can be \todo{interpreted either} as critical and particularly meaningful data or as outliers (i.e. observations that are not generated by the underlying statistical model).}

\subsection*{Application to Outlier Detection} 
Following \cite{chatterjee1986influential} (in the context of linear regression), we argue that the KLD-based measure of influence of an observation can be {also} used for effective outlier detection in data modeled with the HMM. Indeed, if $X_j = x_j$ is an outlier, then it must have a strong influence on the posterior distribution of the hidden variables, which in turn,  must differ significantly from the posterior distribution of the hidden variables conditioned on all the observations but $X_j$. In other words, we expect the KLD distance $K_j$ to be significantly larger when $X_j=x_j$ is an outlier \todo{(an illustration supporting this assumption can be found in the Supplementary Material).} 

\todo{In order to explore whether the KLD is an appropriate measure for outlier detection,}
we considered semi-parametric simulations based on the time series of changes in global temperature described above. The original data is assumed to be free of outliers. 1000 simulations under the null hypothesis H0 (no outliers) were obtained by random sampling $n=106/2=53$ data points in the original time series. 1000 simulations under the alternative hypothesis H1 (presence of outliers) were obtained by sampling $n=53$ data points from the original time series and adding a Gaussian noise $\mathcal{N}(0,\delta^2)$ to each of them with probability $0.05$. Hence, the resulting average number of outliers in each H1 simulation is $0.05\cdot n=2.65$.

\greg{For each simulation, we computed the following global statistics for outlier detection: the maximum $K_j$, the maximum absolute normalized z-score (using a three component mixture model) and the maximum Local Outlier Factor (LOF) score computed with the R package Rlof \cite{Rlof} after rescaling both year and temperature axes.} \todo{Details on the statistics can be found in the Supplementary Material.}

The performances  of the three global statistics for three different values of $\delta$ were assessed with the empirical AUC (computed with \cite{pROC}); the results are depicted in Table \ref{tab:results}. The statistics based on the $Z$-value have very poor performance, whereas the KLD-based statistics has a good discriminating power for $\delta \geq 2.0$. However, the method consisting in computing the LOF scores after normalizing both axes prove to be very performant for each value of $\delta$. All three methods are very fast: it takes less than 0.5 seconds for generating a simulation and computing all three statistics.

\begin{table}[!t] 
\renewcommand{\arraystretch}{1.3}
\caption{Performance of methods for outlier detection: empirical AUC with  $95\%$ confidence intervals, sample size = 400. $\delta$ is the standard deviation of the Gaussian noise characterizing outliers.} 
\label{tab:results} 
\centering 
\begin{tabular}{lccc} 
\hline
Method & $\delta=0.5$ & $\delta=2.0$ & $\delta=3.0$ \\
 \hline
KLD & $0.62\, [0.57,0.68]$ & $0.79\, [0.74,0.84]$ & $0.86\, [0.82, 0.90]$  \\ 
$Z$-value & $0.58\, [0.52,0.64]$ & $0.61\, [0.55,0.66]$ & $0.59\, [0.53,0.65]$  \\
LOF & $0.73\, [0.68, 0.78]$ & $0.93\, [0.90,0.96]$ & $0.94\, [0.91,0.96]$   \\ 
 \hline 
 \end{tabular} 
 \end{table}


\section{Conclusions}

An interesting question in Hidden Markov Models is assessing the relative importance of each observation with respect to the sequence of hidden states. In order to measure how influential is the $j$-th observation, we suggest to use the Kullback-Leibler distance $K_j$ between the conditional distribution of the hidden sequence given the whole observation sequence and the conditional distribution of the hidden sequence given all the observations but the $j$-th one. \greg{\todo{The suggested measure of influence focuses} on the posterior distribution of the hidden sequence rather than on the parameter estimate (like in sensitivity analysis) and it is therefore suitable for problems where the information of interest is the hidden sequence (speech recognition \cite{gold2011speech}, genetics {\cite{li2010mach}}, bioinformatics {\cite{camproux2004hidden}})}

The most important contribution of this letter is a novel linear complexity algorithm for computing the measures of influence of all the observations. Our algorithm is based on simple recursions derived from the forward-backward algorithm for HMMs and can be easily extended in order to take into account pairs, triplets or $h$ consecutive observations. In this case the complexity is $\mathcal{O}(nhm^2)$. The algorithm can be also extended to more complex configurations of observations, the resulting complexity depends on the combinatorics of the configuration.

\greg{We showed that the KLD influence measure can help to detect} outliers in time series modeled by HMMs, the intuition being that anomalies must be more influential than other observations. In this context, the KLD-based method proves to be efficient for global detection, even though it is less performant than specific methods such as the LOF algorithm (after appropriate rescaling). 

\greg{However, the main interest of the KLD measure of influence is the detection of individual observations which, rather than being outliers, are meaningful values playing a critical role in the problem under consideration. New knowledge can be uncovered by investigating the most influential observations found with our influence measure. For example, in the context of protein structure analysis, structural alphabet are encoded through HMMs \cite{camproux2004hidden}. Pointing out highly influential residuals in the encoding through the KLD measure might reveal interesting structural properties (e.g. alternative 3D-structures).}



%




\ifCLASSOPTIONcaptionsoff
  \newpage
\fi



%
%

%

\bibliographystyle{IEEEtran}
\bibliography{biblio}

\begin{thebibliography}{10}
\providecommand{\url}[1]{#1}
\csname url@samestyle\endcsname
\providecommand{\newblock}{\relax}
\providecommand{\bibinfo}[2]{#2}
\providecommand{\BIBentrySTDinterwordspacing}{\spaceskip=0pt\relax}
\providecommand{\BIBentryALTinterwordstretchfactor}{4}
\providecommand{\BIBentryALTinterwordspacing}{\spaceskip=\fontdimen2\font plus
\BIBentryALTinterwordstretchfactor\fontdimen3\font minus
  \fontdimen4\font\relax}
\providecommand{\BIBforeignlanguage}[2]{{%
\expandafter\ifx\csname l@#1\endcsname\relax
\typeout{** WARNING: IEEEtran.bst: No hyphenation pattern has been}%
\typeout{** loaded for the language `#1'. Using the pattern for}%
\typeout{** the default language instead.}%
\else
\language=\csname l@#1\endcsname
\fi
#2}}
\providecommand{\BIBdecl}{\relax}
\BIBdecl

\bibitem{ephraim2002hidden}
Y.~Ephraim and N.~Merhav, ``Hidden {M}arkov processes,'' \emph{IEEE
  Transactions on Information Theory}, vol.~48, no.~6, pp. 1518--1569, 2002.

\bibitem{rabiner1989tutorial}
L.~Rabiner, ``A tutorial on hidden {M}arkov models and selected applications in
  speech recognition,'' \emph{Proceedings of the IEEE}, vol.~77, no.~2, pp.
  257--286, 1989.

\bibitem{gold2011speech}
B.~Gold, N.~Morgan, and D.~Ellis, \emph{Speech and audio signal
  processing}.\hskip 1em plus 0.5em minus 0.4em\relax Wiley Online Library,
  2011.

\bibitem{Durbin1999}
R.~Durbin, S.~R. Eddy, A.~Krogh, and G.~Mitchison, \emph{{Biological Sequence
  Analysis : Probabilistic Models of Proteins and Nucleic Acids}}.\hskip 1em
  plus 0.5em minus 0.4em\relax {Cambridge University Press}, Jul. 1999.

\bibitem{Bishop2006}
C.~M. Bishop, \emph{{P}attern {R}ecognition and {M}achine {L}earning
  ({I}nformation {S}cience and {S}tatistics)}.\hskip 1em plus 0.5em minus
  0.4em\relax Secaucus, NJ, USA: Springer-Verlag New York, Inc., 2006.

\bibitem{do2003fast}
M.~Do, ``Fast approximation of {K}ullback-{L}eibler distance for dependence
  trees and hidden {M}arkov models,'' \emph{IEEE Signal Processing Letters},
  vol.~10, no.~4, pp. 115--118, 2003.

\bibitem{cook1977detection}
D.~Cook, ``{D}etection of {I}nfluential {O}bservations in {L}inear
  {I}nference,'' \emph{Journal of Statistical Planning and Inference}, vol.~37,
  pp. 51--68, 1977.

\bibitem{johnson1985influence}
W.~Johnson, ``Influence measures for logistic regression: {A}nother point of
  view,'' \emph{Biometrika}, vol.~72, no.~1, pp. 59--65, 1985.

\bibitem{fridlyand04}
J.~Fridlyand, A.~Snijders, D.~Pinkel, D.~Albertson, and A.~Jain, ``Hidden
  {M}arkov models approach to the analysis of array {CGH} data,'' \emph{Journal
  of Multivariate Analysis}, vol.~90, no.~1, pp. 132--153, 2004.

\bibitem{camproux2004hidden}
A.~Camproux, R.~Gautier, P.~Tuffery \emph{et~al.}, ``A hidden markov model
  derived structural alphabet for proteins,'' \emph{Journal of molecular
  biology}, vol. 339, no.~3, pp. 591--606, 2004.

\bibitem{li2010mach}
Y.~Li, C.~Willer, J.~Ding, P.~Scheet, and G.~Abecasis, ``{MaCH}: using sequence
  and genotype data to estimate haplotypes and unobserved genotypes,''
  \emph{Genetic epidemiology}, vol.~34, no.~8, pp. 816--834, 2010.

\bibitem{silva2008upper}
J.~Silva and S.~Narayanan, ``{U}pper {B}ound {K}ullback--{L}eibler {D}ivergence
  for {T}ransient {H}idden {M}arkov {M}odels,'' \emph{IEEE Transactions on
  Signal Processing}, vol.~56, no.~9, pp. 4176--4188, 2008.

\bibitem{sahraeian2011novel}
{S.M.E. Sahraeian} and {B.J. Yoon}, ``A novel low-complexity {HMM} similarity
  measure,'' \emph{IEEE Signal Processing Letters}, vol.~18, no.~2, pp. 87--90,
  2011.

\bibitem{data}
R.~J. Hyn­d­man, ``{T}ime {S}eries {D}ata {L}ibrary,''
  \url{http://​data​.is/​T​S​D​Ldemo}, {A}ccessed on September, 24
  2012.

\bibitem{chatterjee1986influential}
S.~Chatterjee and A.~Hadi, ``Influential observations, high leverage points,
  and outliers in linear regression,'' \emph{Statistical Science}, vol.~1,
  no.~3, pp. 379--393, 1986.

\bibitem{Rlof}
\BIBentryALTinterwordspacing
Y.~Hu, W.~Murray, and Y.~Shan, \emph{Rlof: {R} parallel implementation of Local
  Outlier Factor ({LOF})}, 2011, {R} package version 1.0.0. [Online].
  Available: \url{http://CRAN.R-project.org/package=Rlof}
\BIBentrySTDinterwordspacing

\bibitem{pROC}
X.~Robin, N.~Turck, A.~Hainard, N.~Tiberti, F.~Lisacek, J.-C. Sanchez, and
  M.~M\"uller, ``{p{ROC}: an open-source package for R and S+ to analyze and
  compare {ROC} curves},'' \emph{BMC Bioinformatics}, vol.~12, p.~77, 2011.

\bibitem{markou2003novelty}
M.~Markou and S.~Singh, ``Novelty detection: a review, {P}art 1: statistical
  approaches,'' \emph{Signal Processing}, vol.~83, no.~12, pp. 2481--2497,
  2003.

\bibitem{yeung2003host}
D.~Yeung and Y.~Ding, ``Host-based intrusion detection using dynamic and static
  behavioral models,'' \emph{Pattern recognition}, vol.~36, no.~1, pp.
  229--243, 2003.

\bibitem{zhang2005semi}
D.~Zhang, D.~Gatica-Perez, S.~Bengio, and I.~McCowan, ``Semi-supervised adapted
  {HMMs} for unusual event detection,'' in \emph{IEEE Computer Society
  Conference on Computer Vision and Pattern Recognition CVPR 2005},
  vol.~1.\hskip 1em plus 0.5em minus 0.4em\relax IEEE, 2005, pp. 611--618.

\bibitem{shah2006integrating}
S.~Shah, X.~Xuan, R.~DeLeeuw, M.~Khojasteh, W.~Lam, R.~Ng, and K.~Murphy,
  ``Integrating copy number polymorphisms into array {CGH} analysis using a
  robust {HMM},'' \emph{Bioinformatics}, vol.~22, no.~14, pp. e431--e439, 2006.

\bibitem{siu2006robust}
M.~Siu and A.~Chan, ``A robust {V}iterbi algorithm against impulsive noise with
  application to speech recognition,'' \emph{{IEEE} {T}ransactions on {A}udio,
  {S}peech, and {L}anguage {P}rocessing}, vol.~14, no.~6, pp. 2122--2133, 2006.

\bibitem{chatzis2007robust}
S.~Chatzis and T.~Varvarigou, ``A {R}obust to {O}utliers {H}idden {M}arkov
  {M}odel with {A}pplication in {T}ext-{D}ependent {S}peaker
  {I}dentification,'' in \emph{IEEE International Conference on Signal
  Processing and Communications ICSPC 2007}.\hskip 1em plus 0.5em minus
  0.4em\relax IEEE, 2007, pp. 804--807.

\bibitem{chatzis2009robust}
S.~Chatzis, D.~Kosmopoulos, and T.~Varvarigou, ``Robust sequential data
  modeling using an outlier tolerant hidden {M}arkov model,'' \emph{IEEE
  Transactions on Pattern Analysis and Machine Intelligence}, vol.~31, no.~9,
  pp. 1657--1669, 2009.

\bibitem{breunig2000lof}
M.~Breunig, H.~Kriegel, R.~Ng, and J.~Sander, ``{LOF}: identifying
  density-based local outliers,'' in \emph{ACM Sigmod Record}, vol.~29,
  no.~2.\hskip 1em plus 0.5em minus 0.4em\relax ACM, 2000, pp. 93--104.

\end{thebibliography}

%
\vspace{-1.5cm}
\begin{IEEEbiographynophoto}{Vittorio Perduca}
received the M.Sc. and Ph.D. in mathematics from the University of Turin (Italy) and the M.Sc. in computational biology from Paris Descartes University (France) in 2004, 2009 and 2011 respectively.

Dr. Perduca's postdoctoral fellowship is supported by the \textit{Fondation Sciences Math\'ematiques de Paris}; his  research interests include belief propagation algorithms for Bayesian networks with applications in computational biology.
\end{IEEEbiographynophoto}
\vspace{-1.5cm}
\begin{IEEEbiographynophoto}{Gregory Nuel}
received the Ph.D. in mathematics and the \emph{Habilitation \`a Diriger des Recherches} in 2001 and 2007 respectively, both from the University of Evry (France).

Dr. Nuel currently works as a senior researcher for CNRS. His topics of interests include models with incomplete data, Bayesian networks, motifs in random sequences, and a wide range of biomedical applications.
\end{IEEEbiographynophoto}










\vspace{1cm}

\appendix[Supplementary Material with Technical Details]

\section*{Application to Outlier Detection}

\change{In this section, we give more details on our application of the KLD measure of influence to the detection of outliers in HMMs.}

A few methods based on the HMM have been developed for outlier detection \cite{markou2003novelty,yeung2003host,zhang2005semi}. In the main paper we consider a related yet different problem, namely the detection of outliers in data \emph{modeled} with the HMM, for instance time series. \change{This problem is not new in the literature, for instance an \emph{ad hoc} model for outliers in data modeled by HMMs was introduced in a Bayesian framework in~\cite{shah2006integrating}. Other authors suggested to tackle the problem by  means of a robust Viterbi algorithm performing a joint decoding and outlier detection during the Viterbi search \cite{siu2006robust}.} Following \cite{chatterjee1986influential} (in the context of linear regression) we suggest to detect outliers in HMMs by means of our KLD-based measure of observation influence.

An outlier is an observation that is not generated by the underlying statistical model. Since HMMs are intrinsically heterogeneous, the detection of outliers in data modeled by HMMs is a challenging problem. For instance, change point-detection methods based on HMMs are known to be particularly sensitive to the presence of outliers in the sense that a single outlier can result in a segment consisting in just one point \cite{chatzis2007robust,chatzis2009robust}. 

As explained in the main paper, we expect the KLD distance $K_j$ to be significantly larger when $X_j=x_j$ is an outlier. 

\begin{figure}[!t]
\centering
\includegraphics[width=0.45\textwidth]{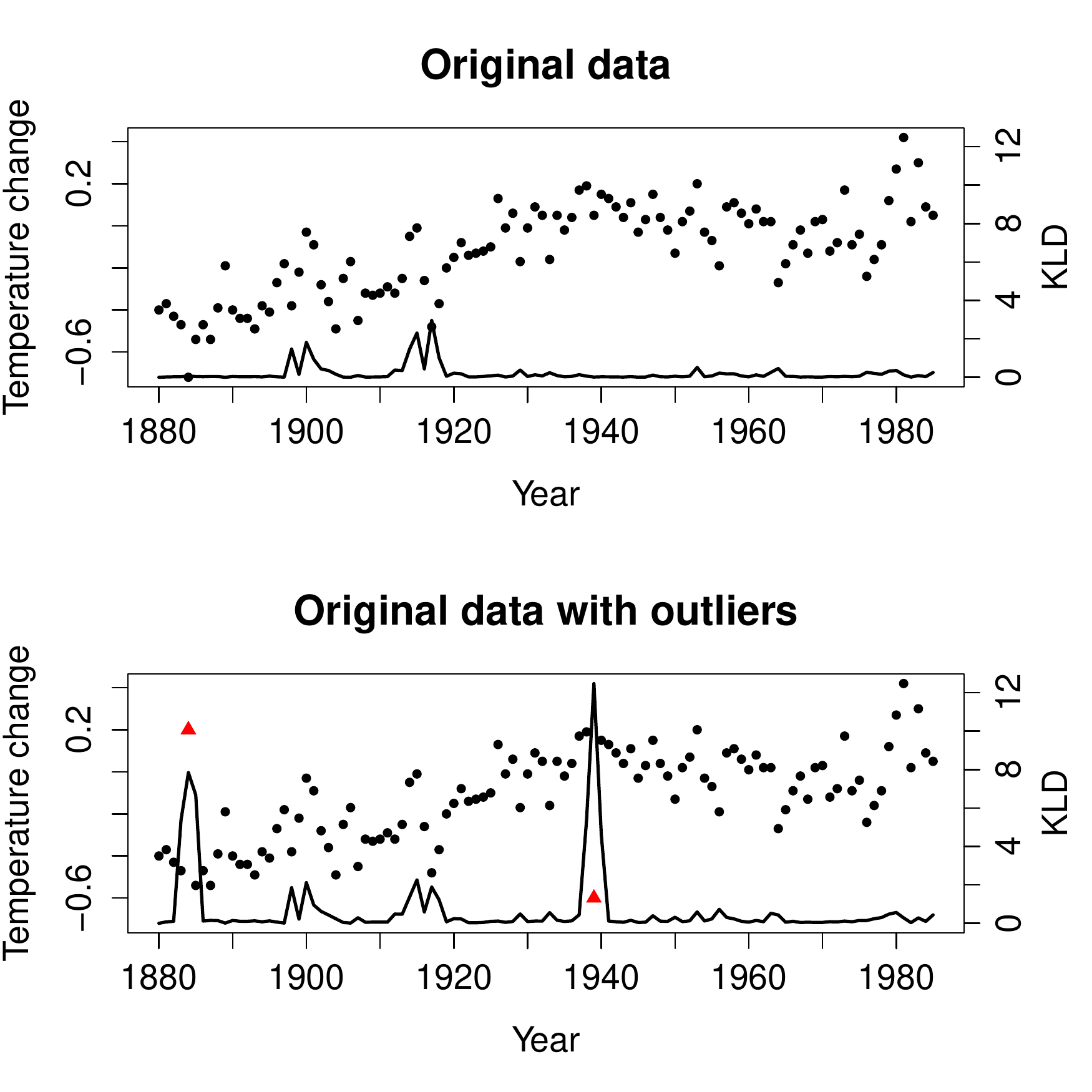}
\caption{KLD function $(K_{j})_{j=1,\ldots,n}$ for the original time series (top) and for the times series with two outliers artificially added (triangles): $x_{1884}=0.2$, $x_{1939}=-0.6$.}
\label{fig:data_out}
\end{figure}

\change{\subsection{Illustration}
Consider the same time series consisting of 106 annual changes in global temperature between 1880 and 1985 as in the main paper. The dataset can be modeled with an HMM in which each observation follows a Gaussian distribution whose mean depends on the corresponding hidden state. The upper plot in Fig.~\ref{fig:data_out} shows the KLD function computed after estimating the parameters with the EM algorithm (same function as in the upper plot of Fig.~2 in the main text but on a different scale). We assume that the original data is outlier free (as we explain in the main text, the peaks in the figure can be interpreted as pointers to meaningful observations). However, if the dataset contains outliers, can we detect them with our measure of influence based on the KLD?  In order to answer this question, we manually added two outliers and re-computed the KLD function, after re-estimating the parameters. The results are depicted in the lower plot of Fig. \ref{fig:data_out} and clearly show that the KLD function has two peaks in the two outliers. } 

\subsection{\change{Comparison with other methods}}

\change{We give here the details of the empirical comparison study whose results are reported in the main paper.}

\textbf{Data.} We considered semi-parametric simulations based on the time series of changes in global temperature described above. The original data is assumed to be free of outliers. 1000 simulations under the null hypothesis H0 (no outliers) were obtained by random sampling $n=106/2=53$ data points in the original time series. 1000 simulations under the alternative hypothesis H1 (presence of outliers) were obtained by sampling $n=53$ data points from the original time series and adding a Gaussian noise $\mathcal{N}(0,\delta^2)$ to each of them with probability $0.05$. Hence, the resulting average number of outliers in each H1 simulation is $0.05\cdot n=2.65$.

We tested the global hypothesis H1 that the data contain \emph{at least one outlier} against the hypothesis H0 that there are no outliers with the following alternative methods: 

\textbf{KLD-based method.} For each simulation $q$ we estimated the parameters in the HMM modeling the dataset with the EM algorithm and then computed the global statistics  
$$
T_q=\max_{j=1,\ldots,n} K_j.
$$

\textbf{Z-value.} For each simulation $q$ we clustered the data with the $k$-means algorithm ($k=3$) and then computed the $Z$-value \todo{$Z_j=\frac{x_j-\mu}{\sigma}$} of each data point $x_j$ with respect to the mean $\mu$ and standard deviation $\sigma$ of its cluster. We considered the global statistics
$$
S_q=\max_{j=1,\ldots,n}|Z_j|.
$$ 


\textbf{Local Outlier Factor (LOF).} The LOF algorithm is a density based method \cite{breunig2000lof}. For each data point, the LOF score is calculated by comparing the local density of the point (defined as the inverse of the average distance from its $r$-nearest neighbors) to the average of the densities of its neighbors. The score is interpreted as a measure of whether the point is in a denser or sparser region of the dataset. A ranking of the points as outliers is obtained by sorting them according to their LOF scores. 


The LOF score depends on the choice of the distance parameter $r$; as suggested in \cite{breunig2000lof} for each point we took the maximal LOF score on a range of integer values for $r$, namely $r\in\{10,\ldots,20\}$.
We considered the global statistics
$$
L_q= \max_{j=1,\ldots,n} \max_{r=10,\ldots,20} LOF_r (\tilde{x}_j,\tilde{t}_j),
$$
where $\tilde{x}_j$ and $\tilde{t}_j$ are the standardized values of $x_j$ and $t_j$ (i.e. we rescaled both axes before computing the LOF scores). The LOF scores were computed using the R package Rlof \cite{Rlof}.


\textbf{ROC analysis.} We assessed the performance of each method by means of the empirical Area Under the Curve (AUC). The AUC measures the surface under the Receiver Operating Characteristic (ROC) curve and can be qualitatively interpreted as follows: $\text{AUC} \leqslant 0.6$ means ``fail''; $0.6 < \text{AUC} \leqslant 0.70$ means ``poor''; $0.7 < \text{AUC} \leqslant 0.80$ means ``fair''; $0.8 < \text{AUC} \leqslant 0.9$ means ``good''; $0.9 < \text{AUC} \leqslant 1.0$ means ``excellent''. AUC computations were performed with the R package pROC \cite{pROC} using the statistics computed for each method and simulation.

\end{document}